# The Concurrent OLSA test: A method for speech recognition in multi-talker situations at fixed SNR

[1,2]Jan Heeren, [2,3]Theresa Nüsse, [4]Matthias Latzel, [2,3]Inga Holube, [1,2,5]Volker Hohmann, [1,2,6]Kirsten Wagener, [2,6]Michael Schulte

[1]HörTech gGmbH, Oldenburg, Germany

[2]Cluster of Excellence Hearing4All, Oldenburg, Germany

[3]Institute of Hearing Technology and Audiology, Jade University of Applied Sciences, Oldenburg, Germany

[4]Sonova AG, Stäfa, Switzerland

[5]Auditory Signal Processing and Cluster of Excellence "Hearing4all," Department of Medical Physics and Acoustics, University of Oldenburg, Oldenburg, Germany

[6]Hörzentrum Oldenburg, Oldenburg, Germany

**Abstract**

A multi-talker paradigm is introduced that uses different attentional processes to adjust speech recognition scores with the goal to conduct measurements at high signal-to-noise ratios. The basic idea is to simulate a group conversation with three talkers and a participant. Talkers alternately speak sentences of the German matrix test OLSA. Each time a sentence begins with the name "Kerstin" (call sign), the participant is addressed and instructed to repeat the last words of all sentences from that talker, until another talker begins a sentence with "Kerstin". The alternation of the talkers is implemented with an adjustable overlap time that causes an overlap between the call sign "Kerstin" and the target words to be repeated. Thus, the two tasks of detecting "Kerstin" and repeating target words are to be processed at the same time as a dual task. The paradigm was tested with 22 young normal-hearing participants for three overlap times (0.6 s, 0.8 s, 1.0 s). Results for these overlap times show significant differences with median target word recognition scores of 88%, 82%, and 77%, respectively (including call sign and dual task effects). A comparison of the dual task with the corresponding single tasks suggests that the observed effects reflect an increased cognitive load.

**Introduction**

In daily life, hearing-impaired people have difficulties understanding speech in multi-talker situations (Bronkhorst, 2000; Humes et al., 2006; Singh et al., 2010; Desjardins, 2011), even though the signal-to-noise ratios (SNR) are mostly above speech reception thresholds (SRT; Smeds et al., 2012). To assess, e.g., benefits of hearing devices or performance differences between participant groups in these situations, multi-talker speech recognition tests are needed that are sensitive for variations in speech recognition in the targeted level and SNR range. Such tests are sensitive if the participants recognize some, but not all, of the target words, otherwise there are floor or ceiling effects. At fixed SNRs, there are basically four approaches to adjust the sensitivity when designing such a test: 1. selecting appropriate background noises (Bronkhorst, 2000), 2. manipulating the speech signals (e.g. Schlueter et al., 2015; Oxenham & Simonson, 2009), 3. using reverberation (Westermann & Buchholz, 2015; Rennies et al., 2011), and 4. increasing cognitive load (e.g. Desjardins, 2011; Singh et al., 2010; Humes et al., 2006). For a flexible test paradigm, that is applicable to different acoustical scenes as background noises and uses natural speech signals, the first three possibilities are rejected. Instead, the developed method, the Concurrent OLSA (CCOLSA), aims at increasing cognitive load to adjust speech recognition for the German matrix test (Oldenburg sentence test, OLSA, Wagener, Brand & Kollmeier, 1999).

Shin-Cunningham & Best (2008) published a review on the role of attention in cocktail-party situations. They point out that normal-hearing listeners have no difficulties understanding speech in cocktail-party situations due to their ability for selective attention, which is defined as the enhancement of the neural representation of an auditory object (e.g. a talker) while the allocation of resources on competing objects is reduced (Shin-Cunningham & Best, 2008). These resources are understood as the complete auditory system including cognitive stages and the auditory periphery. Considering that the resources are limited, the complementary concept of divided attention can be defined as concurrent processing of two or more auditory objects in distinct channels (Allport et al., 1972). Consequently, when listening to speech-in-speech with selective attention, the recognition of attended speech is increased while unattended speech is ignored to some degree. Still, unattended speech is semantically processed (Aydelott et al., 2015; Bentin et al., 1995) and thus influences recognition of attended speech by blocking resources (Schneider et al., 2007). Recognizing simultaneous speech with divided attention is also possible, but it is limited to a few words (for two talkers: 2-3 words each) and depends on the background noise (Best et al., 2010; Meister et al., 2018). The maximum number of words that can be recognized simultaneously is approximately three (Ericsson et al., 2004), as at least it is possible to detect a targeted word (call-sign) in a mixture of three by up to 55 % correct responses. This score was observed for three talkers with different voice characteristics including different sex, whereas increasing the similarity of voices decreases the recognition scores (Brungart, 2001; Ericsson et al., 2004). These call-sign experiments were based on the Coordinate Response Measure paradigm (CRM, Bolia et al., 2000), that allows for investigations of speech recognition in multi-talker settings combined with call-sign based attention guidance, which means that the target talker was unknown until it was discernable through a call sign.

In the CRM, sentences of the format "Ready (call sign), go to (color) (number) now." are presented simultaneously for a specified number of different talkers. Subjects have to detect a call sign (e.g., "baron") and identify the color-number combination (e.g., "white seven") being said by the talker who spoke the call sign. Humes et al. (2006) used a two-talker setting and tested elderly hearing-impaired (EHI) and young normal-hearing participants (YNH). Both groups of participants yielded correct-response ratios close to 100% in the easiest condition (dichotic presentation of the two sentences, different gender of the two talkers, selective attention listening with prior visual announcement of target talker). Discarding monaural listening conditions, the strongest effects were observed for a divided attention condition (approx. 25%-points decrease in recognition score) and for call-sign based target-talker detection (approx. 10%-points decrease). The divided attention condition was realized by taking responses for a trial after giving announcements for the next trial that had to be memorized. Meister, Rählmann & Walger (2018) used a similar effect to design a speech recognition test paradigm at high SNR (+2 or +6 dB). For two simultaneously presented sentences in noise (male and female voice), the effect of background noise was tested with respect to the word position, while the sentences of the female voice were to be repeated first. The recognition scores for all words of the prioritized sentence (female talker) were above 90%, while the secondary sentence (male talker) showed recognition scores of below 10%, except for the last word of the sentence. The recognition score for this last word was around 40% depending on the level of the background noise, being a measure for cognitive load.

In group conversations, listening with selective attention might be useful, but as talkers take turns, this listening mode gets interrupted. Brungart & Simpson (2007) investigated turn-taking using the CRM approach with target position switches for several talker configurations. They tested switches between target directions, target talkers, and both combined. The main finding was that any switches were followed by short-time decreases in speech recognition for configurations with 3-4 talkers. Comparable results were observed by Meister et al. (2020), who used a similar paradigm based on three target talker positions (-60°, 0°, and 60°) and the OLSA. These findings (Brungart & Simpson, 2007; Meister et al., 2020) were measured trial-by-trial with pauses to take the participants' responses. A closer look at timing effects in such switches was taken by Best et al. (2008), who found that switching targets for fixed talker positions has higher switching costs (i.e., decrease in recognition due to the target switch) if the sentences are cued with pauses <250 ms than for longer pauses. Another timing aspect was found by Müller et al. (2018). Their data suggests that increasing the speech rate increases the cognitive load. Considering that also unattended speech is semantically processed (Aydelott et al., 2015; Bentin et al., 1995), this may be relevant for both target and concurrent speech.

The cited publications on turn-taking are based on the presentation of synchronous speech tokens. In those experiments, the task is detecting a single target call sign against similar stimuli, while the distractors can be discarded. Although, turn-taking in real-life situations is mostly not synchronous. Aiming for a more realistic paradigm, it can be considered that an interrupter typically competes with the current target speech. Thus, there are two sources with possibly equal priorities (target talker and interrupter).

The concept of interrupting talkers was used to design a novel multi-talker speech test, for which call signs compete with target speech. Based on the presentation of OLSA sentences from three alternating talkers, alternation was shifted into interruption by applying defined temporal overlaps between sentences from the different talkers. Furthermore, the total speech rate depends on this overlap, while the speed of each sentence stays constant. This has an additional decreasing effect on speech recognition scores (Müller et al., 2008). Consequently, following hypotheses have to be proven to validate this design:

1. The concurrence of target words and call signs leads to a decrease in speech recognition scores.

2. The overlap time between sentences can be used to further adjust the decrease in speech recognition scores.

To investigate hypothesis 1, a single-/dual-task comparison was carried out for call sign detection and speech recognition reflecting the influence of the attention mode. Hypothesis 2 was assessed by a comparison of three overlap times (0.6 s, 0.8 s, and 1 s).

**Method**

**Participants**

22 young normal-hearing participants (eleven females) were measured. Their age ranged from 19-27 years with a mean of 22 years. All participants were unexperienced regarding the OLSA and showed pure-tone thresholds of ≤ 25 dB HL at all audiometric test frequencies between 0.125 and 8 kHz. The pure-tone-average at 0.5, 1, 2, and 4 kHz of the better ear (PTA4) was -1.5 dB HL (SD: 3.1 dB HL). All participants gave their informed consent prior to inclusion in the study and received a reimbursement of 12 € per hour. The experiment was approved by the ethics committee ("Kommission für Forschungsfolgenabschätzung und Ethik") of Carl von Ossietzky University in Oldenburg, Germany (Drs. 34/2017).

## Stimuli

The developed method CCOLSA is based on the German matrix test OLSA. Within the OLSA, each sentence consists of five words following the structure name-verb-number word-adjective-noun. For each of these word classes, the matrix contains ten words. OLSA speech recordings of three talkers were used in this study: the standard male OLSA talker (M1, Wagener et al., 1999), the standard female OLSA talker (F1, Wagener et al., 2004), and a second male talker (M2) from the multi-lingual OLSA (Hochmuth et al., 2014). Initially, speech levels for the three corpora were calibrated to 68 dB SPL. To achieve equal intelligibility of the three corpora, speech recognition thresholds (SNR for a speech recognition score of 50%) were measured for all of them in a pre-test (method: word-scoring based 1 up-2 down procedure with adaptive speech level; Brand & Kollmeier, 2002). Differences in the mean SRTs were compensated by increasing the presentation level of talker M2 to 69.0 dB SPL and the level of talker F1 to 69.2 dB SPL for the main experiment. The speech was presented against a diffuse cafeteria noise (Grimm & Hohmann, 2019), which was also presented at 68 dB SPL. It was recorded in the cafeteria of the University of Oldenburg, campus Wechloy, in ambisonics B-format (first order). The character is highly diffuse and stationary with smaller, rather distant, fluctuating components.

## Setup

*Loudspeaker array*

Measurements were conducted in a lecture room with some acoustical optimizations (acoustic ceiling and curtains, moveable damping items) at Jade University of Applied Sciences in Oldenburg. The room has a volume of 262 m$^3$ with a T30 of 0.46 s. In the room, a loudspeaker setup was realized as shown in Figure 1. Signals were presented via ten loudspeakers (Rokit 5, KRK Systems, Deerfield Beach, USA) positioned on a circle with a radius of 1.5 m. For the presentation of the diffuse background, eight loudspeakers were set up equally spaced on the circle. Via the frontal speaker, also the female speech was presented. Additionally, two loudspeakers were placed at -60° and 60° azimuth for the presentation of the two male talkers. Participants were seated in the center of the loudspeaker setup on a twistable office chair without armrests. The presentation of stimuli was controlled using the virtual acoustics toolbox TASCAR (Grimm, Luberadzka, & Hohmann, 2019). Speech was presented using nearest speaker panning and the cafeteria recording (recorded in 1st order ambisonics) was upsampled and presented in 3rd order ambisonics. The CCOLSA test was implemented in MATLAB 2014 (MathWorks Inc., Natick, USA)

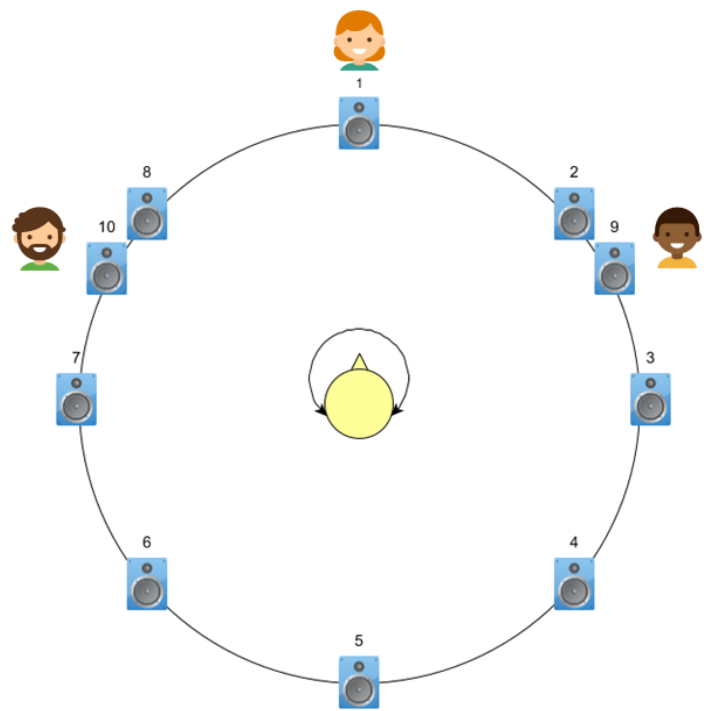

Figure 1: Sketch of the loudspeaker setup with a participant seated in the center and positions of the talkers.

### Head tracking

The head orientation of the participants was monitored with a wireless head-tracking system based on a Bosch BNO055 integrated measurement unit (Bosch Sensortec GmbH, Reutlingen, Germany). The sensor was combined with a Teensy-LC microcontroller (PJRC, Sherwood, USA) and a HC-05 Bluetooth module. This set was boxed and attached to the participants' heads using a cap. The head orientation data was buffered in the LabStreamingLayer (https://github.com/sccn/labstreaminglayer/) and recorded in TASCAR (Grimm, Luberadzka, & Hohmann, 2019) to store it with a matched time scale. In a preprocessing stage, the head position data was smoothed (2-s-average). During the analysis of the dual task (see below), the head orientation at the time of a target sentence presentation (2 s after the sentence onset) is considered. As participants do not necessarily turn their heads by the full amount of +- 60° to face the loudspeaker of the male target talkers, the orientation was counted successful if the 2-s-average exceeded a threshold of +-25°.

## Paradigm

### General design

The CCOLSA paradigm follows the general idea of measuring speech recognition in multi-talker situations with talkers that call for the participant's attention. This was achieved by three talkers (M1 at -60°, F1 at 0°, and M2 at +60°) that speak OLSA sentences with a defined overlap between sentences from different talkers. One of the ten occurring names in the OLSA corpus, the name "Kerstin", was defined to be the call sign. In Figure 2 a sketch of an exemplary sequence of sentences is given.

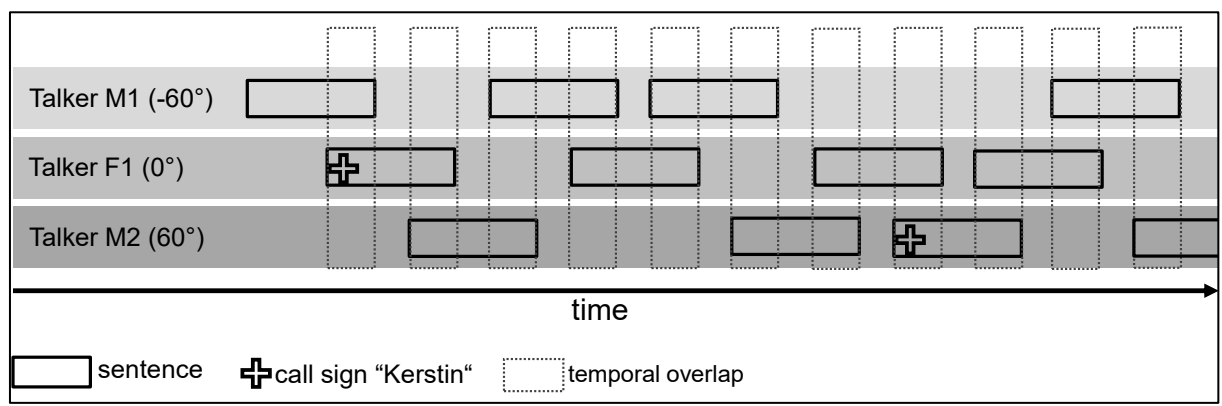

*Figure 2: Exemplary sequence of sentences; sentences were timed with a defined temporal overlap and were presented randomly from the three talkers; some sentences started with the name "Kerstin", which was defined to be the call sign.*

*Tasks*

Three tasks were performed in this scenario addressing three different listening modes:

Single task 1 (call sign detection): "Name the number of the correct loudspeaker or point a finger in the direction if the word "Kerstin" is presented". In this single task, participants have to permanently monitor all talkers in order not to miss the first word of any sentence, because the call sign "Kerstin" might occur. The task was performed for 10 call signs, occurring every 10-13 s. The test was introduced by an initial phase of "non-Kerstin sentences" from random talkers with a duration of approximately 20 s.

Single task 2 (speech recognition for a fixed talker): "Repeat the last words of all sentences for a fixed talker." During this speech recognition task, participants focus on the target talker, which does not change during the measurement. Sentences from the other two talkers can be ignored. In this task 2, OLSA test lists of 20 sentences were presented for the three talkers in separate measurements. The experimenter announced the target talker position before starting the measurement. During an initial phase of approx. 20 s, only the two non-target talkers presented sentences, so that the first sentence of the previously announced target talker was recognizable as the first target sentence without the need for a call sign. After completion of one test list, the new target talker was announced and the next speech recognition measurement started.

Dual task: "Repeat the last words of all sentences from the talker who said "Kerstin" the latest. Always turn your body on a rotatable chair so that you face the talker you attend to." After a 20 s-initial phase of sentences from random talkers, the first sentence starting with the call sign "Kerstin" occurred (call sign sentence). This indicated the first target talker for which participants had to repeat the last words of all sentences (target sentences) until "Kerstin" occurred from another talker and the target talker changes. The dual task requires simultaneous attention to the last word of the current target sentence (target word) and the first word of the following sentence (potentially a call sign), which appear at the same time (see Figure 3). The tested overlap times were chosen so that the target word is always completely overlapping with the beginning of the following sentence from the next talker and there are never more than two talkers active. A dual task measurement contained 21 call sign sentences (seven per talker), each of which followed by 1-3 target sentences and one fake target sentence. In total, 60 target sentences were presented (20 per talker). To achieve a constantly high difficulty, all target talker changes were introduced by a fake target sentence (S3 in Figure 3) from the previous target talker, meaning that a call sign sentence was never presented next to a distractor sentence. Fake target sentences were not evaluated in the analysis of the results.

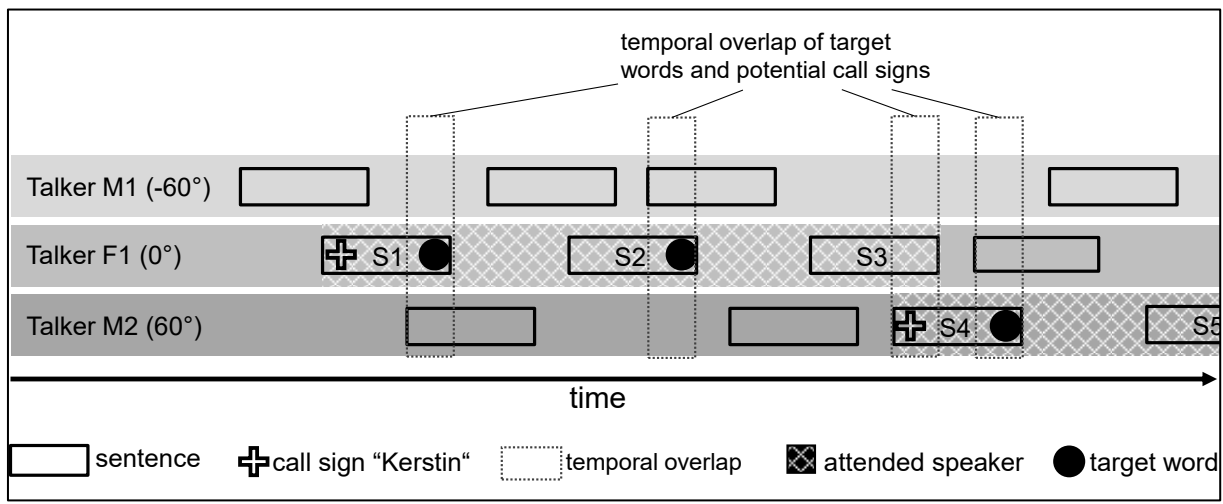

*Figure 3: Attention pattern in the dual task for an exemplary sequence of sentences. Participants have to repeat the last words of all sentences for the talker, who spoke a sentence starting with the name "Kerstin" (call sign) the latest; target words and call signs occur at the same time due to a defined overlap time between sentences across talkers.*

Responses for all tasks were given orally by the participants during the ongoing presentation and were entered by the experimenter by pressing a button on a touch screen. The sentence presentations were controlled continually by the software, to adapt to the individual response times of the participants without presentation pauses. Typically, two or three distractor sentences (sentences from the non-target-talkers) were presented during a response. Directly after a response, the next target sentence was presented. This design resulted in a quick target-and-response flow, where target words occurred every 5-6 s. Due to this flow, the experimenter was able to judge if the participant could not respond, immediately. When this was the case, the experimenter was instructed to quickly push the button for "target not recognized". The resulting button-press flow avoided indirect feedback by changes in the behavior of the experimenter, as if it would have occurred if misses would have been judged by the software, e.g. by a count down.

*Outcome variables*

Outcomes of the two single tasks are the percentage of correct responses for call sign detection and for speech recognition. In the dual task, the participants do not directly respond on the call signs (e.g. by specifying the number of the loudspeaker). To allow for an analysis of the call sign detection during dual task, two criteria were considered: 1. correct response of target words between two call signs, and 2. the head orientation of the participant measured with the head tracker. If one of these criteria was met, a call sign was counted as successfully detected. The impact of the two factors (responses and head tracking) on the results was analyzed separately.

The speech recognition scores in the dual task were calculated as percent of correctly repeated target words and is given in the results as "speech recognition incl. missed-call-sign (CS) effects". To distinguish between effects of missed call signs and failure in target word recognition, a correction for misses caused by call sign detection errors was applied based on the two criteria given above. Therefore, the "net speech recognition" score was calculated as percent of correctly repeated target words while taking only target words for successfully detected call signs into account.

## Conditions and procedure

The CCOLSA paradigm was tested for three overlap times of the sentences: 0.6 s, 0.8 s, and 1.0 s (see Figure 2 and 3). Per overlap condition, the single tasks, i.e., call sign detection and speech recognition for fixed target talkers, were performed. Additionally, the dual task including recognition of last words for changing target talkers was conducted. Each task took approx. 5-6 min. In total, nine measurement tasks and two training tasks were conducted per participant. The procedure was split into two sessions of 2 h per participant that were conducted within three weeks. In the first session, a general anamnesis, a pure-tone audiometry and the single tasks (including a short training phase) were conducted. Call sign detection was tested first and speech recognition (fixed talker) second. Additionally, two neurophysiological tests were executed. The second session started after participants had performed another neuropsychological test sequence (approx. 20 min). Then, a longer training including the two single and the dual task at 0.6 s overlap and the data collection for the dual task at the three overlaps were conducted, followed by another 20 min of neurophysiological tests. The overlap conditions were tested in randomized order for all tasks in both sessions. The analysis of the neurophysiological tests is beyond the scope of this contribution and is addressed in (Nuesse et al., submitted).

## Results

### Single and dual task

Figure 4 shows a boxplot of the results (medians, interquartile ranges, and whiskers) in the CCOLSA dual task and the two corresponding single tasks for the three overlap times of 0.6, 0.8, and 1.0 s. The left panel shows call sign detection scores, the right panel shows target word recognition scores. Target word recognition scores for the dual task are given as a net value, i.e. considering only those target words with correct call sign detection, and additionally as a total score including the missed-call-sign effects. Most of the data are not normally distributed (Shapiro-Wilk test, $\alpha = 0.05$). Therefore, statistical analyses were conducted using non-parametric statistical tests.

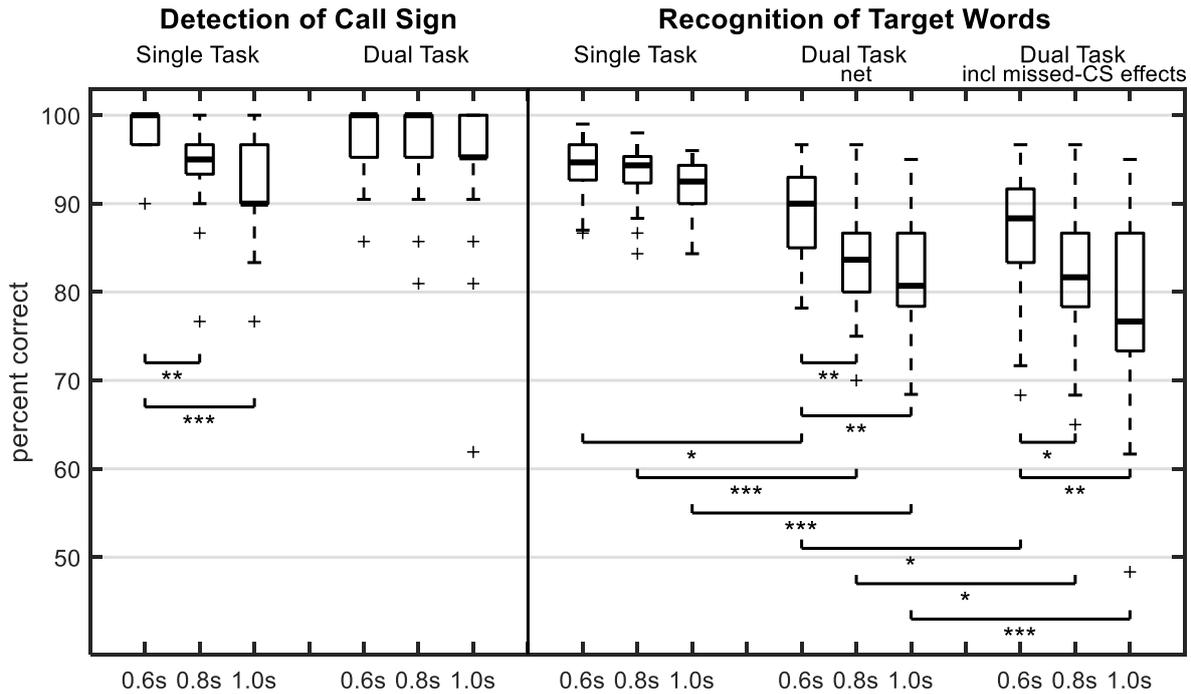

*Figure 4: Boxplot of the call sign detection scores and target word recognition scores measured for the CCOLSA dual task and the corresponding single tasks for three overlap times. In the net values of the dual task target word recognition, only those target words with correct call-sign detection are considered. Statistical significance is indicated by asterisks (Wilcoxon test; * p<0.05, ** p<0.01, *** p<0.001).*

Participants detected the call sign with a median score of ≥90% correct in both the single and the dual task condition (see Figure 4). No significant differences between the tasks were found. The median score in target word recognition was 93-95% for the three overlaps in the single task. In the dual task, this score significantly dropped to 81-90% (net scores) depicting a difference in speech recognition of 5-12%-points. Significance was tested using a Wilcoxon test (Bonferroni corrected; $p_{0.6s}=0.015$, $r_{0.6s}=0.60$, $T_{0.6s}=213$; $p_{0.8s}<0.001$, $r_{0.8s}=0.86$, $T_{0.8s}=250$, $p_{1.0s}<0.001$, $r_{1.0s}=0.81$, $T_{1.0s}=244$). The median total target word recognition scores including the effect of missed call signs range from 77-88% and are significantly lower than the net scores and depicting a difference in speech recognition in the single task of 7-16%-points (Wilcoxon test, Bonferroni corrected; $p_{0.6s}=0.035$, $r_{0.6s}=0.54$, $T_{0.6s}=36$; $p_{0.8s}=0.054$, $r_{0.8s}=0.51$, $T_{0.8s}=28$, $p_{1.0s}=0.004$, $r_{1.0s}=0.68$, $T_{1.0s}=91$).

## Overlap dependency

The results for the three different overlap conditions show a dependency of the correct-response score from the overlap time of the sentences, i.e., increasing the overlap time decreases the correct-response score (Figure 4). Significance was tested using Friedman tests (Bonferroni-corrected) and post-hoc Wilcoxon tests (Bonferroni-corrected) for the Friedman tests with significant results (see Table 1 and Figure 4). This effect of overlap duration was significant for target word recognition in the dual task (net and including missed call-sign effects) and call sign detection in the single task.

*Table 1: Results of analysis of performance differences between overlaps with the Friedman test and post-hoc comparisons using Wilcoxon test, bold fond indicates statistically significant values; p-values are Bonferroni-corrected to account for multiple testing.*

| Friedman test | | Detection of call sign | | Recognition of target words | | |
|---|---|---|---|---|---|---|
| | | Single task | Dual Task | Single task | Dual task, net | Dual task, incl. missed-CS effects |
| $\chi^2$ | | **25.77** | 3.38 | 8.02 | **18.70** | **13.00** |
| $p$ | | **< 0.001** | 0.553 | 0.054 | **< 0.001** | **0.005** |
| Paired comparison between Overlaps / s | Wilcoxon test | | | | | |
| 0.8 vs. 0.6 | $T$ | **4.5** | | | **36.0** | **34.0** |
| | $p$ | **0.002** | | | **0.010** | **0.013** |
| | $r$ | **0.738** | | | **0.627** | **0.606** |
| 1.0 vs. 0.6 | $T$ | **0.0** | | | **20.0** | **27.5** |
| | $p$ | **< 0.001** | | | **0.002** | **0.004** |
| | $r$ | **0.800** | | | **0.737** | **0.686** |
| 1.0 vs. 0.8 | $T$ | 28.0 | | | 73.5 | 61.0 |
| | $p$ | 0.198 | | | 0.719 | 0.299 |
| | $r$ | 0.392 | | | 0.251 | 0.351 |

## Head orientation in dual task

Call sign detection scores in the dual task were calculated either indirectly by reviewing the target word responses after a call sign (if a correct response was given, the call must have been detected), or by additionally evaluating the participants' head orientation recorded using a head tracker. In the dual task call sign detection results the influence of the head orientation is not visible, because the scores are near to 100% either way.

Nevertheless, as one missed call sign causes 2-4 missed target words, there is a significant difference between speech recognition scores incl. missed call-sign effects and net scores. Figure 5 shows the net target word recognition scores determined considering or discarding head orientation of the head tracker data. The condition with consideration of head orientation is identical to the net scores in Figure 4. The condition discarding head orientation incorporated only the first criterion (correct response of target words between two call signs) when calculating net scores. Discarding head orientation, the median values of percent correct scores range from 86-91%, instead of 81-90% if head orientation is considered. The effect of the overlap time is also statistically significant when head orientation is discarded (Friedman test, Bonferroni corrected, $\chi^2$=9.2, p=0.03). In post-hoc tests one significant paired comparison was found: 0.6s vs. 1.0s (Wilcoxon test, Bonferroni corrected, p=0.013, r=0.61, T=39). However, the difference caused by the head-tracker data is significant for all overlap times (Wilcoxon test, Bonferroni corrected; $p_{0.6s}$=0.043, $r_{0.6s}$=0.43, $T_{0.6s}$=0; $p_{0.8s}$=0.001, $r_{0.8s}$=0.68, $T_{0.8s}$=0, $p_{1.0s}$=0.003, $r_{1.0s}$=0.63, $T_{1.0s}$=0).

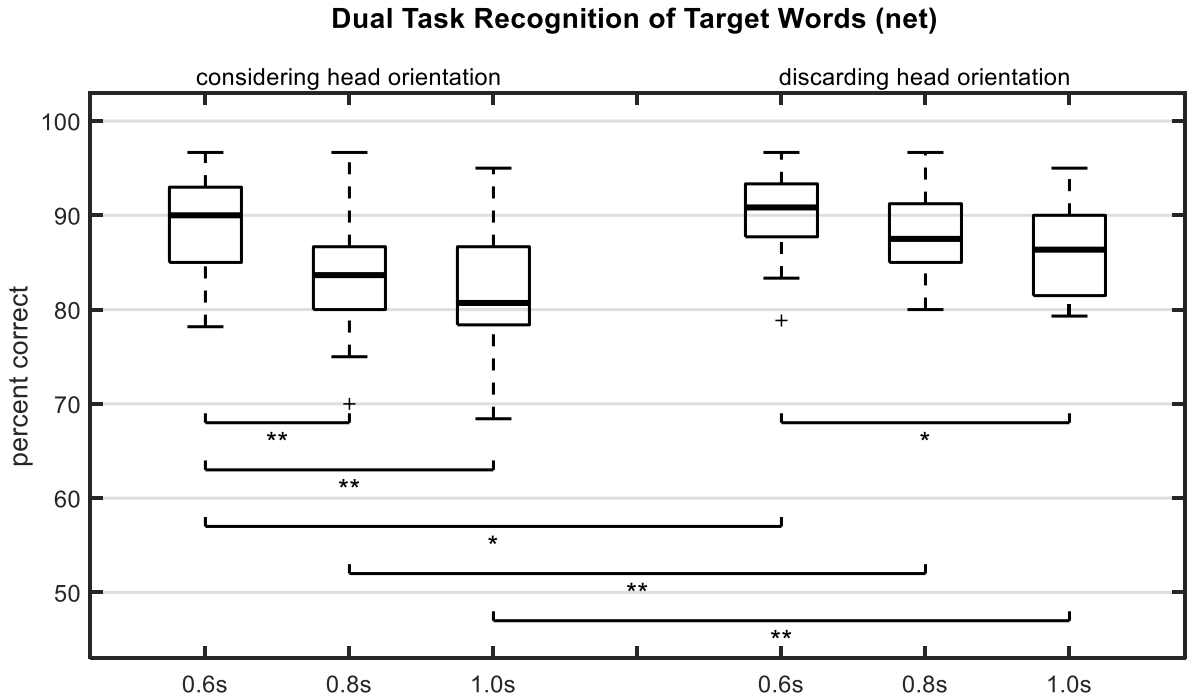

*Figure 5: Influence of head tracker data on the compensation of target word recognition scores for call sign detection errors; boxes 1-3 are identical with boxes 10-12 in Figure 4; statistical significance is indicated by asterisks (Wilcoxon test; * p<0.05, ** p<0.01).*

## Discussion

### Single Tasks

Speech recognition in the single task condition was expected to be close to 100% (Humes et al., 2006; Bronkhorst, 2000), as the measurements were conducted at approx. 0 dB SNR with two active talkers at a time. This expectation was confirmed for the shortest overlap time. Higher overlap times seemed to result in a decrease in target word recognition scores, but this effect was not significant. The results for call sign detection in the single task condition meet the expectation that participants are able to perform it almost perfectly (Ericson et al., 2004; Hawley et al., 2004). Nevertheless, this condition shows a significant effect of the overlap time. Since increasing the overlap time increases the amount of words to be recognized as potential call signs, the cognitive load might be enlarged (Müller et al., 2018). Furthermore, switches of the spatial position of the target speech results in decreasing speech recognition (Brungart and Simpson, 2007). As known from binaural sluggishness experiments (Culling and Summerfield, 1998), changing the binaural focus takes around 100 ms. After each sentence onset, the participant's attention is shortly triggered to the current talker to verify whether "Kerstin" has been said. Afterwards, the attention is immediately needed to monitor the other two talkers and recognize the next onset. This serial focusing and unfocusing process demands a high level of concentration and is presumably related to cognitive load.

## Dual Task

In the dual task, the target word recognition scores are significantly smaller than for the single task. This confirms the hypothesis that the concurrence of target words and call signs leads to a decrease in speech recognition scores (hypothesis 1). The net target word recognition (corrected for call sign detection errors) is 5-12%-points lower than for the single task condition. As this decrease reflects the dual-task costs compared to selective attention listening, it is comparable with the decrease by 10%-points in CRM results caused by call-sign based target announcement (Humes et al., 2006). The overall recognition score in the dual task (recognition of target words including missed-CS effects) is additionally decreased by 2-4%-points and indicates that the detection of call signs also slightly suffers in the dual task, although the direct comparison of single and dual task for call sign detection shows no significant difference. This reflects that missing a single call sign leads to 2-4 missed target words. The observed differences between the overlap conditions confirm the hypothesis that the speech recognition rates can be adjusted by the overlap time (hypothesis 2).

## Advantages of the CCOLSA

Compared to other approaches (e.g. Bolia et al.,2000; Meister et al., 2018; Meister et al., 2020), the CCOLSA test focuses on the comparison of attentional processes in an ecologically valid listening situation including background noise as well as speech distractors. The CCOLSA test has the advantage that the difficulty of the test can be adjusted by the overlap time. In future applications, test difficulty can potentially be adjusted to the target group, e.g., elderly people or hearing aid users. Additionally, the CCOLSA test allows for testing speech recognition combined with directional effects. This combination might be explicitly useful to test spatial noise reduction algorithms in hearing devices, where trade-offs between sound quality and spatial cue preservation are a common issue. Furthermore, a single-/dual-task comparison can be applied to investigate factors as discussed here.

**Technical Issues and Limitations**

For matrix tests as the OLSA, training effects are known and were extensively examined for the original male and a synthetic female talker (Schlueter et al., 2016; Nuesse et al., 2019). In this study, an explicit training phase with speech recognition measurements of 60 training sentences as suggested in the literature was not included. Instead, shorter training phases of every specific task were conducted. The measurements always started with the detection of call signs in single task for which no data about training effects is available yet. After the single-task detection of call signs, speech recognition in single task was assessed. Because of the familiarization with the test material during the detection task and the high SNR leading to speech recognition scores near 100 % (above 90 % in median for all overlaps), the impact of training effects is estimated to be low in the CCOLSA. Nevertheless, it cannot be ruled out that participants went through various training effects in the different tasks, including the effects due to speech material and familiarization to the different talkers as well as to the spatial situation.

Furthermore, measurements were conducted in an acoustically treated, yet reverberant, lecture room. The influence of room acoustics is expected to be small but could not be excluded. To evaluate the CCOLSA setup in other settings, comparative measurements in a multi-center study might help to estimate the influence of room acoustics.

## Conclusion

The CCOLSA paradigm was introduced to measure speech recognition in multi-talker situations based on different types of attentional processes. The OLSA corpus was combined with the concept of call sign tests for three talkers and a novel timing control, so that call signs and target words occur with a defined and overlapping timing. The difficulty of the test can be adjusted by the overlap time between call sign and target word, which allows for applicability at high SNRs. Under the same test paradigm, a call sign detection test, a speech recognition test and a dual task can be performed. As the talkers are spatially separated, the test can for example be applied to the evaluation of beamformers in hearing devices or any other algorithms influencing the binaural listening conditions.

## Acknowledgments

Funded by Deutsche Forschungsgemeinschaft (DFG, German Research Foundation) Project-ID 352015383, SFB 1330 B1 and C4. We would like to thank Sven Kissner and Uwe Simmer for their support with the head-tracking system as well as Jule Pohlhausen and Mira Richts for data collection.